\documentclass[aps,showpacs]{revtex4}
\usepackage{graphics}
\usepackage{graphicx}
\usepackage{latexsym}
\newcommand{\beqa}{\begin{eqnarray}}
\newcommand{\eeqa}{\end{eqnarray}}

\begin{document}
\title{Effects of Resonant and Continuum States on the Neutrino-Nucleus Cross Section.}
\vskip2cm
\author{O. Civitarese$^1$\footnote{Corresponding author \\ e-mail: osvaldo.civitarese@fisica.unlp.edu.ar}, R. J. Liotta$^2$ and M. E. Mosquera$^3$}
\affiliation{{$^1$\small\it Dept of Physics, University of La
Plata}  {\small\it c.c.~67 1900, La Plata, Argentina}}
\affiliation{{$^2$\small\it KTH, Albanova University Center,
SE-10691 Stockholm, Sweden}} \affiliation{{$^3$\small\it Faculty of
Astronomy and Geophysics, University of La Plata, La Plata,
Argentina}}
\date{\today}

\begin{abstract}
Estimates of the neutrino-nucleus cross section, for the
charged-current process $\nu+^{208}\rm{Pb} \rightarrow
e^-+^{208}\rm{Bi}^*$, are presented. The nuclear structure
calculations have been performed by considering bound, resonant,
and continuum states in the single-particle basis used to
construct correlated proton-particle neutron-hole configurations.
The observed features of the spectrum of $^{208}$Bi have been
reproduced, as accurately as possible, by diagonalizing  a
phenomenological multipole-multipole interaction. Calculations of
the cross section, for values of q $\leq 200 $ MeV, were
performed, and the dependence of the results upon the choice of
the residual proton-neutron interaction was investigated. It is
found that the inclusion of resonant states in the calculation of
the nuclear wave functions increases the neutrino-nucleus cross
section, and that the contribution of the continuum is negligible.
\end{abstract}


\maketitle

key words: neutrino-nucleus cross section, nuclear structure,
resonant states.

\section{Introduction}

During the last decade an intense effort, both theoretical and
experimental, was devoted to elucidate neutrino properties and the
connection between them and the physics of supernova
\cite{ref1,ref2,ref3}. The phenomena of neutrino oscillations was
confirmed \cite{sno,sk}, and with it, the solution of the solar
neutrino problem \cite{ref5,ref6} was found. This is a cornerstone
upon which we base our present knowledge of neutrino properties,
and it will certainly be followed by other breakthroughs in other
neutrino-related fields, like nuclear double-beta-decay \cite{dbd}
and neutrino astrophysics \cite{nastro}. The prospect of detecting
supernova-neutrino-flavor oscillations by analyzing the response
of various nuclei to neutral- and charged-current interactions,
was advanced by Fuller, Haxton and McLaughlin \cite{fuller}, and
later pursued by Elliot \cite{elliot} and Engel, McLaughlin and
Volpe \cite{engel}. In these processes an electron-neutrino
interacts inelastically with a target nucleus, leaving it in an
excited state (neutral-current interactions), or it is absorbed
and changes a neutron into a proton, thus resulting in an outgoing
electron and a residual nucleus with charge $Z_f=Z+1$
(charged-currents). For the case of charged-current interactions,
that is the $(\nu_x, e^-)$ channels, the theoretical estimate of
the total inelastic cross section varies from $10^{-40}$ cm$^2$,
in the case of the process $^{23}$Na$(\nu_e, e^-)^{23}$Mg, to
$10^{-38}$ cm$^2 $ for the case of $^{208}$Pb$(\nu_e,
e^-)^{208}$Bi \cite{fuller}.

The lepton sector of the reaction is described in terms of a
single-flavor neutrino $(\nu_e)$ to which we may add mixing terms
due to neutrino oscillations \cite{engel}. The nuclear structure
sector is governed by the strength distribution of the complete
set of multipole excitations induced by the energy-momentum
transferred from the lepton sector to the hadronic sector. A
realistic description of the neutrino-nucleus cross section
requires a fairly detailed knowledge of the nuclear spectrum
\cite{elliot,fukugita,kuramoto}, a goal which may be achieved by a
direct diagonalization of the realistic nuclear interaction or by
approximate methods like the Tamm-Dancoff Approximation (TDA) or
the Random Phase Approximation (RPA). An alternative to the use of
individual nuclear states is the use of energy-weighted sum rules
and, subsequently, the replacement of the detailed
nuclear-energy-level distribution by a few energy-centroids which
concentrate all of the intensity carried by each multipolarity
\cite{kuramoto}. Though the low energy sector of the nuclear
spectrum may be determined experimentally, thus allowing for a
detailed comparison with theoretical predictions, the high energy
part of it may be unreachable by standard spectroscopic methods
and therefore it is not so well established theoretically, from
the nuclear structure point of view. However, future measurements
on neutrino-nucleus interactions may improve our knowledge on this
high energy regime, as suggested by Volpe \cite{volpe1}.

Lead perchlorate was suggested, as a detector of choice
\cite{elliot} because the estimated neutrino-nucleus cross section
may be one or two orders of magnitude larger than the one
corresponding to the scattering of neutrinos by $^{23}$Na. Also,
future experimental efforts at SNOLAB, like HALO, will be based on
neutrino reactions on Pb \cite{noble}.  However, in the case of
lead, one encounters additional complications caused by the
description of the spectrum of a double-odd-mass heavy-nucleus,
like $^{208}$Bi. In addition to the relatively well known strength
distribution of the Isobaric Analog State (IAS) and Gamow-Teller
(GT) resonances one should consider forbidden transitions leading
to unnatural parity states. From the microscopic point of view one
needs to calculate the eigenstates of the nuclear Hamiltonian
belonging to complete sets of angular momentum $J$ and parity
$\pi$. It is evident that an exact shell model diagonalization can
not be performed for a heavy-mass nucleus in a large model space.
Thus, one may resort to approximations like weak-coupling schemes
or TDA (RPA) treatments \cite{bmvII}. Because the energy deposited
in the nucleus by the neutrino may be of the order of few tens to
few hundreds of MeV, one may expect that the contribution of
nuclear states in the continuum should also be considered, and,
therefore, be added to the contributions resulting from the low
energy region of the nuclear spectrum \cite{fukugita}. However,
continuum RPA calculations find a serious discrepancy between
measured and calculated cross sections for neutrino induced
reactions \cite{hayes,hayes-towner}.

The role of neutrino induced reactions on lead an iron was
discussed in a series of RPA calculations \cite{kolbe}. The
formalism of \cite{kolbe} takes into account the correct momentum
dependence of the operators entering the definition of the
charge-current interactions, a fact leading to a reduction of the
calculated values as compared with calculations performed at
$q=0$. The calculations of the various channels entering inclusive
neutrino-nucleus interactions, of \cite{kolbe,langanke}, yield
values of the cross section, for lead targets, which are dominated
by neutron-emission processes from $^{208}$Bi. This feature, which
emerges from the results of \cite{kolbe} corresponding to LSND
pion-decay in-flight neutrino energies and supernovae neutrinos,
indicates that the contribution of the neutron emission channel is
typically few orders of magnitude larger than the calculated
$\gamma$ and proton-emission channels. The contributions coming
from proton emission from $^{208}$Bi are significantly smaller, of
the order of $10^{-41}$cm$^2$, than the neutron-emission channel,
of $10^{-38}$cm$^2$. The contribution of the $\gamma$ emission
from $^{208}$Bi amounts also to a fraction of the cross section,
of $10^{-39}$cm$^2$. The smallness of the contribution of the
proton-emission channel is particularly interesting, since one may
expect that the proton emission from isovector resonances (mostly
Isobaric Analogue and Gamow-Teller states) may be significant
\cite{ian}.

In dealing with the theoretical description of processes involving
the continuum part of nuclear spectra one has to distinguish among
the various approaches that have been used for this purpose. One
can thus mention: 1) the analytical extension of the discrete
portion of the spectrum [18]; ii) the continuum shell model
\cite{vol03} and iii) the shell model in the complex energy plane
(CXSM) \cite{id02}. The CXSM has been used extensively since its
first application nearly 20 years ago \cite{ver88}. It will also
be applied in this paper. A brief description of its main
ingredients will be given in the next Section.

In the context of the neutrino-nucleus scattering, the nuclear
structure component of the calculation will then be centered upon
the theoretical prediction of the distribution of intensities for
the nuclear transitions induced by the multipole operators of the
leptonic current. Because the energy transferred to the nucleus may
be large, of the order of 100-200 MeV, the probability to excite
nuclear states with a large component on a single-particle resonance
is also very large. This fact then opens the possibility of finding
enhancement of the cross section, similarly to the one found in
proton-emission  and in cluster-emission.

In this work we have calculated neutrino-nucleus cross sections on
lead, leading to $^{208}$Bi, including resonant and continuum
states in the single-particle basis. In so doing we focus our
attention on the possible effects due to the inclusion of the
continuum. For the sake of the present calculation, we aim at a
quantitative estimate of the contributions due to the continuum,
by performing nuclear structure calculations in Berggren´s
representation \cite{ver88,b68,lio96}. As we shall show, the
effects associated to the inclusion of the nuclear continuum are
extremely small. Contrary to this, the inclusion of
single-particle resonances increases the cross section
significantly.

The formalism of the neutrino-nucleus interactions is briefly
review in Section \ref{formalism}. Details about the nuclear
structure calculations are presented in Section \ref{structure}.
The results of the calculations are discussed in
Section \ref{results}. Finally, the conclusions are drawn in
Section \ref{conclusions}. Details of the theoretical formulations
are presented in the Appendixes \ref{operators}, \ref{density},
and \ref{formfactors}.

\section{Formalism}

In this section we will present the essentials of the
formalism, which includes two main components: a) the treatment of
neutrino-nucleus interaction, and, b) the use of nuclear models to
calculate the participant nuclear wave functions. We shall
focus on the process
\begin{equation}
\nu+^{208}\rm{Pb} \rightarrow e^-+^{208}\rm{Bi}^*,
\end{equation}
thus we have to calculate explicitly an electroweak process, where
the incoming neutrino $\nu$ decays into an electron-W boson-pair,
followed by the nuclear conversion of a neutron into a proton and
the absorption of the $W^+$ boson in the target nucleus, leading
to excited state of the final nucleus $^{208}\rm{Bi}$.

\subsection{Charged-Current Neutrino-Nucleus
Interactions} \label{formalism}

The cross section for the inelastic neutrino-nucleus interaction
in the charged current channel is written
\begin{eqnarray}\label{cross}
\sigma&=&\left( 2 \pi \right)^4 \sum_f \int {\rm d}^3 p_l \delta
\left(E_l +E_f -E_\nu -E_i\right)\left|<l\left(p_l\right);f
\left|H_{eff}\right|\nu\left(p_\nu\right);i>\right|^2 ,
\end{eqnarray}
where $|\nu\left(p_\nu\right);i>$ is the initial product state of
the incoming neutrino $\nu$, with momentum $p_\nu$, and the ground
state of the target nucleus $^{208}$Pb, $\rm{E}_i$ is the energy
of the ground state of $^{208}$Pb, $|l\left(p_l\right);f>$ is the
product state of the outgoing lepton $l$ $(e^-)$ of momentum $p_l$
and the excited state $f$ belonging to the complete set of states
of $^{208}$Bi, with angular momentum $J_f$, parity $\pi_f$ and
energy $\rm{E}_f$. In the convention which we have adopted the
energy $\rm{E}_f$ is measured respect to the ground state of
$^{208}$Pb, $\rm{E}_l$ is the energy of the outgoing lepton
(electron), $\rm{E}_\nu$ is the energy of the incoming neutrino,
and $H_{eff}$ is the electroweak interaction. After separation of
the leptonic and hadronic components of the current-current
interaction, $H_{eff}$, one gets \cite{fukugita}
\begin{eqnarray}
\label{sigma} \sigma&=&\frac{{\rm{G}}^2}{\pi}\cos^2\theta_C \sum_f
p_l E_l F\left(Z_f, E_l \right) \frac{1}{2} \int_{-1}^1 {\rm
d}\left( \cos \theta\right) M_{\rm{Nuc}} .
\end{eqnarray}
The elements of this equation are the electroweak coupling
constant $\rm{G}$, the Cabibbo angle $\theta_C$, the energy and
momentum of the outgoing lepton $(E_l,p_l)$, the Fermi function
$F(Z_f,E_l)$ \cite{report}, and the nuclear transition probability
$M_{\rm{Nuc}}$. The sum runs over the complete set of nuclear
states $|f>$ of $^{208}$Bi, and the variable of integration is the
angle between the incoming neutrino and the outgoing lepton. By
expanding the hadronic weak current in powers of the inverse
nucleon-mass $M_N^{-1}$, and keeping the non-relativistic limit of
small momenta, that is $p_N/M_N << 1$, where $p_N$ is the momentum of
the nucleon, one obtains the standard operators which induce
momentum-dependent axial-vector, vector, weak-magnetic and
pseudoscalar transitions \cite{bmvII}. The nuclear transition
probability $M_{\rm{Nuc}}$ may, therefore, be decomposed into
allowed Fermi and Gamow-Teller transitions, and forbidden and
allowed transitions of higher multipolarities.

The explicit expressions of the components of the nuclear
transition probability $M_{\rm{Nuc}}$ are
\begin{eqnarray}\label{mtau}
M(\tau)=\left|<f | \tau_- e^{i q r} \right|i>  |^2 &=& \frac{4
\pi}{2 J_i+1} \sum_l \left|<J^{\pi}_f || \sum_k \tau_-(k) i^l
j_l\left(q r_k\right) Y_l\left({\bf \hat{r}}_k\right)|| J^{\pi}_i
> \right|^2 ,
\end{eqnarray}
for the isospin dependent operators, and
\begin{eqnarray}\label{msigmatau}
M(\sigma \tau)=\left|<f | \sigma \tau_- e^{i q r}  |i> \right|^2
&=& \frac{4 \pi}{2 J_i+1} \sum_{l,K} \left|<J^{\pi}_f | | \sum_k
\tau_-(k) i^l j_l\left(q r_k\right) \left[ Y_l\left({\bf
\hat{r}}_k\right) \times {\bf \sigma}(k)\right]^{(K)}|| J^{\pi}_i
> \right|^2 ,
\end{eqnarray}
for the spin-isospin dependent operators. To these terms one adds
the moment
\begin{eqnarray}\label{msigmalambda}
M(\Lambda)&=&\left(\frac{5}{6}\right)^2
\frac{4\pi}{(2J_1+1)}\sum_{l,l',K}(-1)^{l/2-l'/2+K}\sqrt{(2l+1)(2l'+1)}
\left(
\begin{array}{ccc}
l&l'&2 \\ 0&0&0
\end{array}
\right) \left\{
\begin{array}{ccc}
1&1&2 \\ l'&l&K
\end{array}
\right\}\nonumber \\
&&\times <J^{\pi}_f | | \sum_k \tau_-(k) i^l j_l\left(q r_k\right)
\left[ Y_l\left({\bf \hat{r}}_k\right) \times {\bf
\sigma}(k)\right]^{(K)}|| J^{\pi}_i
>\nonumber \\
&&\times <J^{\pi}_f | | \sum_{k'} \tau_-(k') i^{l'} j_{l'}\left(q
r_{k'}\right) \left[ Y_{l'}\left({\bf \hat{r}}_{k'}\right) \times
{\bf \sigma}(k')\right]^{(K)}|| J^{\pi}_i
>^*
\end{eqnarray}
The total nuclear matrix element of Eq.(\ref{sigma}) includes
weak-magnetic and pseudoscalar contributions (see \cite{kuramoto}
for details), and it is written
\begin{eqnarray}\label{mnuc}
M_{\rm{Nuc}}&=&\lambda_{\tau} M(\tau)+\lambda_{\sigma \tau} M(\sigma
\tau)+\lambda_\Lambda M(\Lambda)
\end{eqnarray}
The quantities $\lambda_{\tau}$, $\lambda_{\sigma \tau}$, and
$\lambda_\Lambda$ are functions of the momentum and direction of the
outgoing electron and of the nucleon form factors. Their explicit
expressions are given in Refs. \cite{fukugita} and \cite{kuramoto}
and in Appendix \ref{formfactors}. In the above equations
$q=p_l-p_\nu$ is the momentum transferred from the leptonic to the
nuclear sectors, and $\tau_-$ transforms a neutron into a proton.
The transitions induced by the multipole operators of
Eqs.(\ref{mtau}) and (\ref{msigmatau}) obey the standard selection
rules of the conservation of total angular momentum for
parity-changing operators (generally speaking forbidden transitions)
and parity conserving operators (allowed transitions).

\subsection{Nuclear Structure Calculations with Resonant and Continuum
States} \label{structure}

We write the wave function of the excited k-th member of the
$J^{\pi}$ multiplet in $^{208}$Bi as the superposition of
particle(proton)-hole(neutron) states
\begin{equation}
\mid JM,k>=\sum_{pn}C^{(k)}(pn,J^\pi)\mid p n^{-1};JM>.
\end{equation}
and determine the amplitudes $C^{(k)}(pn,J^\pi)$ by a
direct diagonalization of the residual two-body interaction in
the proton-particle-neutron-hole space. The neutron-hole states
are bound-states but the proton-particle states may be
bound-, resonant- or continuum-states.

The single-particle basis which includes all of these
possibilities is  an extension of the conventional single-particle
basis and its use in nuclear structure calculations was advocated
long ago \cite{ver88}. The calculation of
nuclear wave functions and transition densities (Appendix
\ref{density}) in this basis constitutes a mayor difference with
respect to previous calculations
\cite{fuller,elliot,engel,fukugita,kuramoto}, where only bound
states have been included in the single-particle basis. The
expressions of the matrix elements of the two-body residual
interaction, which we have chosen as to reproduce the spectrum of
$^{208}$Bi, are given in the Appendix \ref{density}. Hereafter we
shall review briefly the concepts and general aspects of the
definition of the single-particle with bound, resonant and
continuum states.
This single-particle basis, which forms the Berggren representation,
has been described before e. g. in Refs. \cite{ver88,b68,lio96}.
We will give here only a brief summary of the formalism.

The regular solutions of the Schr\"odinger equation with outgoing
boundary conditions corresponding to a particle moving in a
central potential provide the single-particle bound states and
complex states. The complex states may or may not have physical meaning
but they (as well as the bound state) are poles of the single-particle Green-function. Since
at large distances they behave as $e^{ikr}$ in the complex k-plane
one can uniquely classify them in four categories, namely:

1) bound states, for which Re(k)=0, Im(k)$>$0,

2) anti-bound states, for which Re(k)=0, Im(k)$<$0.

3) outgoing (decay) states for which Re(k)$>$0, Im(k)$<$0,

4) incoming (capture) states for which Re(k)$<$0, Im(k)$<$0.

One sees that only bound states do not diverge at large distances.
One may therefore conclude that only the bound states are
physically meaningful. However, if the wave functions
corresponding to the complex poles are localized, they either
correspond to resonances or anti-bound states which can be
observed or which can produce observable effects \cite{id05}. We
will analyze this feature with some detail in the Applications
below.

In a pioneering paper \cite{b68} Berggren obtained an expansion of
the Green- and  $\delta$-functions in  terms  of the poles of the
Green-function  plus  an  integral along a continuum path in the
complex energy plane, i. e.

\begin{equation}\label{eq:del}
\delta(r-r^\prime)=\sum_n
 w_n(r)w_n(r^\prime) + \int_{L^+} dE u(r,E)u(r^\prime,E)
\end{equation}

The summation runs over all bound states and poles of the Green
function enclosed by the real E-axis and the contour $L^+$. One
can choose quite general forms for the contour, as can be seen in
Ref. \cite{bl}, but it has to finish at infinite on the real
energy axis.
However, as in any shell-model calculation, one cuts the energies
at a certain maximum value. As a direct illustration of these
concepts we shall refer the reader to
Ref. \cite{id05} particularly for details concerning the
integration contour $L^+$.

In Eq. (\ref{eq:del}) the scattering
functions on the contour are denoted by $u(r,E)$ while the wave
functions of the bound single-particle states and the Gamow
resonances are denoted by $w_n(r)$.

An important feature in Eq. (\ref{eq:del}) is that the scalar
product is defined as the integral of the  wave function times
itself, and not its complex conjugate. This is in agreement with
the Hilbert metric on the real energy axis since for bound states or
for scattering states on this axis one can choose the phases
such that the wave functions are real quantities. The prolongation
of the integrand to the complex energy plane, which is done by
applying the Cauchy theorem, allows one to use the same form for
the scalar product everywhere. This metric (Berggren metric)
produces complex probabilities, as has been discussed in detail in
e. g. Ref. \cite{vlm}. Here it is worthwhile to point out that for
narrow resonances such probabilities become virtually real
quantities.

The integral in Eq.(\ref{eq:del}) can be discretize such that

\begin{equation}\label{eq:dis}
\int_{L^+} dE u(r,E)u(r^\prime,E)= \sum_p h_p u(r,E_p)
u(r^\prime,E_p)
\end{equation}
where $E_p$ and $h_p$ are defined by the procedure one uses to
perform the integration. In the Gaussian method $E_p$ are the
Gaussian points and $h_p$ the corresponding weights. Therefore the
orthonormal (in the Berggren metric) basis vectors $\vert
\varphi_j\rangle$ are given by the set of bound and Gamow states, i
. e. $\langle r \vert \varphi_n\rangle=\{w_n(r,E_n)\}$ and the
discretize scattering states, i. e. $\langle r \vert
\varphi_p\rangle=\{\sqrt{h_p}u(r,E_p)\}$. This defines the Berggren
representation.

\section{Results and Discussions}\label{results}

\subsection{Nuclear Structure of Bi}

\begin{figure}[h!]
\includegraphics[height=8cm,width=6cm]{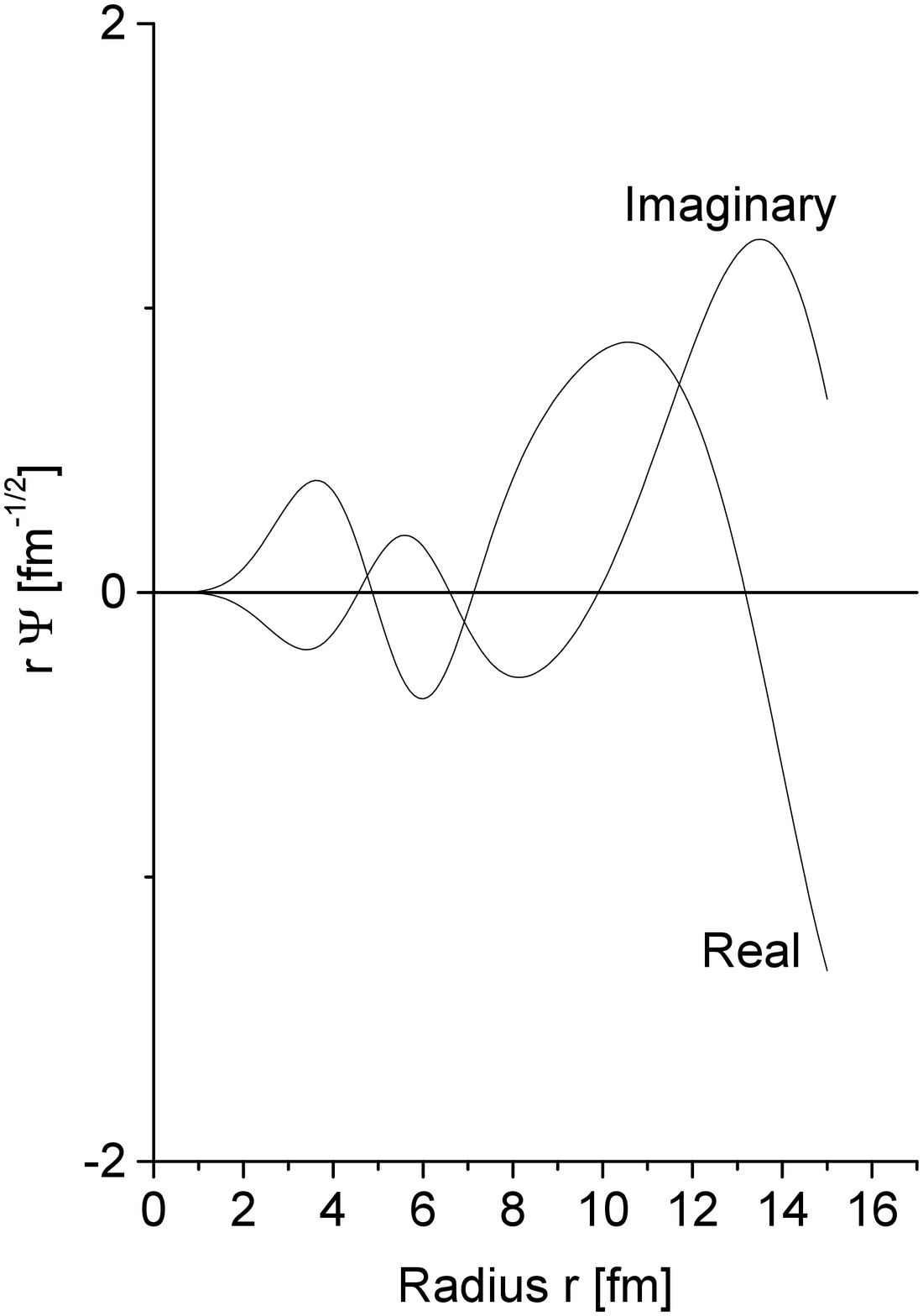}
\caption{Real and imaginary parts of the radial wave function (times $r$) corresponding to a proton state $g_{9/2}$
lying at (17.838,-3.546) MeV, i. e. deep in the continuum.
\label{fig:fig1}}
\end{figure}

The first step in the present calculations is the construction of
the single-particle basis. For this we have
diagonalized the Woods-Saxon plus Coulomb potential to which we
have added a spin-orbit term. The parameter of the single-particle
hamiltonian have been taken from \cite{cvl}. The calculation of the single-particle states corresponding to
all poles were performed by using the method and computer codes of Ref. \cite{ixa}. In order to
illustrate the typical values of the real and imaginary parts of the energies corresponding to
the proton states thus evaluated we present in
Table 1 some selected cases. The
actual single particle basis extends over 136 proton states, with
5 bound states, 4 quasi-bound states, 10 narrow resonant states
and 117 continuum states. We chose the contour containing the
resonances such that they have physical meaning, i. e. that they
are localized inside the nucleus. In other words, the proton is
trapped by the Coulomb and centrifugal barriers inside
the nucleus and therefore the corresponding wavefunction should
also be concentrated inside the nucleus. This wavefunction looks  like the wavefunction
corresponding to a bound state (it is practically a real function) up to large values of the radius. This large
distance depends upon the high of the barrier. The higher the barrier the larger the distance. Beyond it the
wavefunction (including its imaginary part) starts to diverge.
Thus, our contour does not include poles of the
Green function which are very far from the real energy axis. Such
poles cannot be considered resonances but rather they are a part
of the continuum background. Their contribution (as well as the contribution of any physical resonance
which might be left outside the contour) will be taken into account by the scattering states in the contour. An example of a
non-resonant state is the $g_{9/2}$ pole at (17.838,-3.546) MeV shown in
Fig. \ref{fig:fig1}.
One sees that this wave function is  small inside the nucleus
and that it starts to diverge just outside the nuclear surface,
which in this case is located at $r \approx $ 7.5 fm. The
imaginary part is as large as the real part. This is not a
physical state but a part of the continuum background.

The neutron basis includes 16 single-particle bound states bellow
the N$=126$ shell-closure. With these states we have, as a next
step, constructed unperturbed particle-hole states for
configurations with total angular momentum $J$ and parity $\pi$
($J^\pi \leq 10^{\pm}$), and diagonalized the interaction (given
in \ref{vsep}) in the proton (particle)-neutron (hole) states. The
couplings have been adjusted to reproduce the first excited state
for each multipolarity. We have further verified that the dominant
configurations have coefficients similar to those of \cite{alexa}.
We have taken the comparison with the results of \cite{alexa} as a
consistency test of our single-particle basis.
\begin{table}[h!]
\begin{tabular}{cccc}
\hline $lj$&E (real) [MeV]& E (imag) [MeV] & \\ \hline
$h_{9/2}$ & -3.784 & 0 & bound \\
$f_{7/2}$ & -3.541 &0& bound \\
$i_{13/2}$&-1.844 &0& bound \\
$p_{3/2}$&-0.690 &0& bound \\
$f_{5/2}$ & -0.518 &0& bound \\ \hline
$p_{1/2}$&0.491 &0& quasi-bound \\
$g_{9/2}$ & 4.028 &0& quasi-bound \\
$i_{11/2}$&5.434 &0& quasi-bound \\
$j_{15/2}$ &5.960 &0& quasi-bound \\ \hline
$d_{5/2}$ &6.748 & -0.002 & resonant \\
$s_{1/2}$&7.843 & -0.037 & resonant \\
$g_{7/2}$ & 8.087 & -0.001 & resonant \\
$d_{3/2}$& 8.530 & -0.028 & resonant \\
$f_{7/2}$& 12.748 &-0.652 & resonant \\
$h_{11/2}$& 11.390 & -0.022 & resonant \\
$k_{17/2}$& 14.066 & -0.001 & resonant \\
$h_{9/2}$& 15.964 & -0.393 & resonant \\
$j_{13/2}$& 15.086 & -0.005 & resonant \\
$i_{13/2}$&18.143 & -0.575 & resonant \\ \hline
\end{tabular}
\caption{ Proton bound, quasi-bound and resonant states, above the
closure Z$=82$. The values given in columns are the real and
imaginary parts of the energy, in units of MeV. }\label{table1}
\end{table}
%
Table \ref{table2} shows the calculated values of the centroids
corresponding to the set of $0^+$ and $1^+$ states included in the
calculations. Each set of states exhausts the Ikeda's sum rule at
the $1-4\%$ level (real part), given a further indication about
the rightness of the approach concerning the combined effects of
bound, resonant and continuum states.
\begin{table}[h!]
\begin{center}
\begin{tabular}{|c|c|c|}
\hline
$J^\pi$ & Energy (Real) [MeV]& Energy (Imaginary) [MeV]\\
\hline
$0^+$ & 15.21 & -0.125\\
$1^+$ & 16.07 & -0.171\\
\hline
\end{tabular}
\caption{Energy-centroid for $J^\pi=0^+,1^+$ excitations in
$^{208}$Bi. The calculated real and imaginary part of the
energy-centroid, for each multipolarity, are given in the
table.}\label{table2}
\end{center}
\end{table}

The calculated energy-difference between the centroids for pure
Gamow-Teller and Fermi transitions is of the order of 0.86 MeV, a
value which compares rather well with the experimental splitting
between the corresponding GT and IAS resonances, which for A=208
is of the order of 0.5 MeV. The calculated position of the GT
resonance is $\rm{E}_{\rm {GTR}}$=15.6 MeV, thus the calculated
energy difference between the GTR and the IAS is of the order 0.4
MeV, again agreeing with the corresponding experimental value
within experimental limits. Concerning the total
intensity, we have verified that the real part of the Ikeda sum
rule is, for each case, much larger than the imaginary part. The
ratio between the imaginary and real parts of the calculated Ikeda
sum rule is of 1.5 \% (for the GT transitions) and 3.5 \% (for the
Fermi transitions).
This imaginary part can be interpreted as the uncertainty related to
the interference between the resonances and the background \cite{Ber78}.

\subsection{The $(\nu.e^-)$cross section}

After having introduced the features of bound, quasi-bound,
resonant and continuum states we shall present and discuss our
results for the cross section (\ref{sigma}). In performing our
calculations, and analyzing the results, we have focussed on the
following aspects:

a)Dependence of the results upon the chosen residual
proton-neutron interaction

\begin{figure}[h!]
\includegraphics[width=10cm]{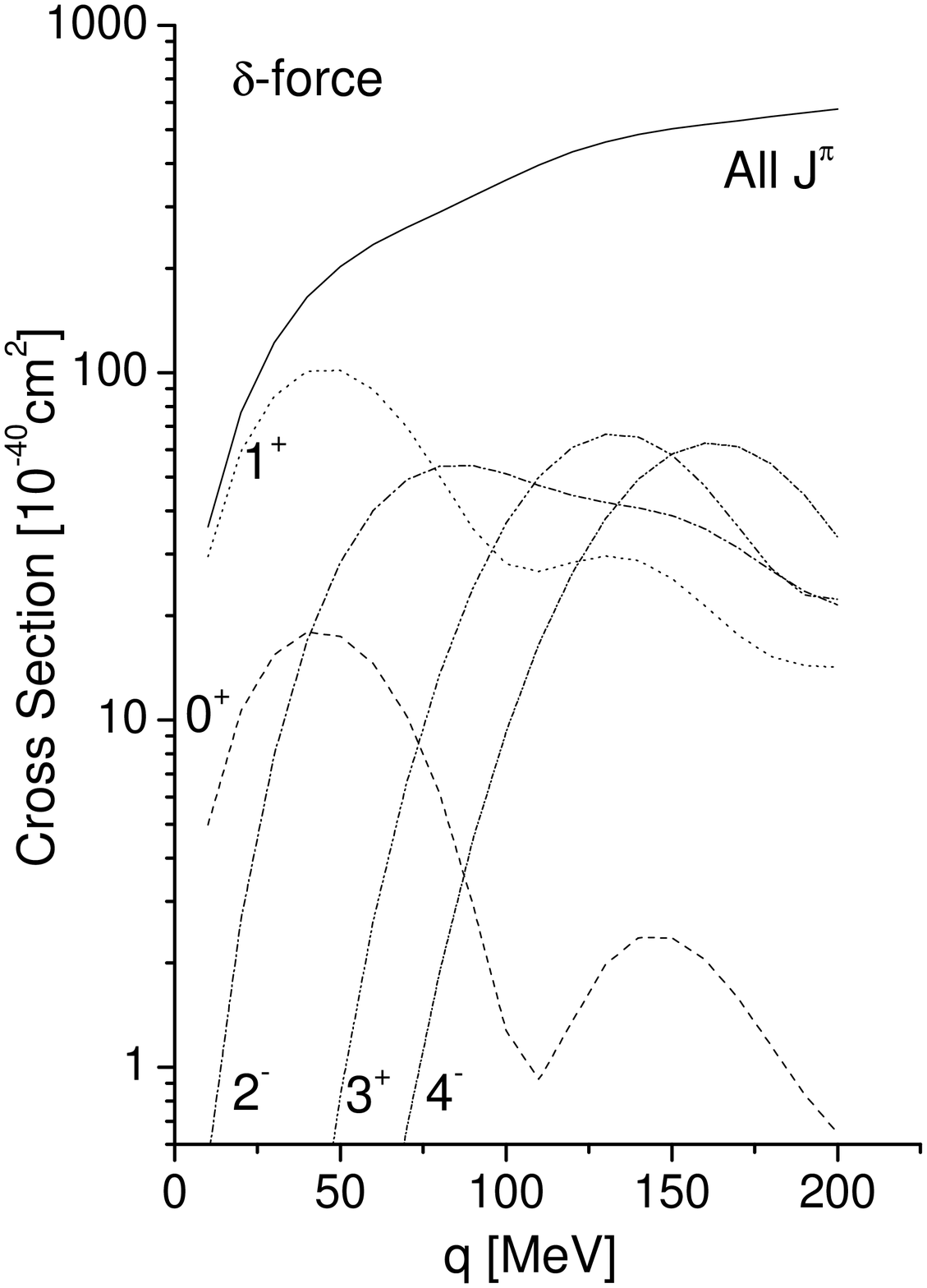}
\caption{Cross section, as a function of the momentum transferred
by the neutrino. The results corresponding to nuclear structure
calculations of states $J^\pi$ in $^{208}$Bi performed in an
ordinary single particle basis and using a $\delta$ force
interaction are shown. Solid line represents the results obtained
by adding up all states up to $J^\pi \leq 10^{\pm}$, the other
curves show the results of some selected states. \label{fig:fig2}}
\end{figure}

\begin{figure}[h!]
\includegraphics[width=10cm]{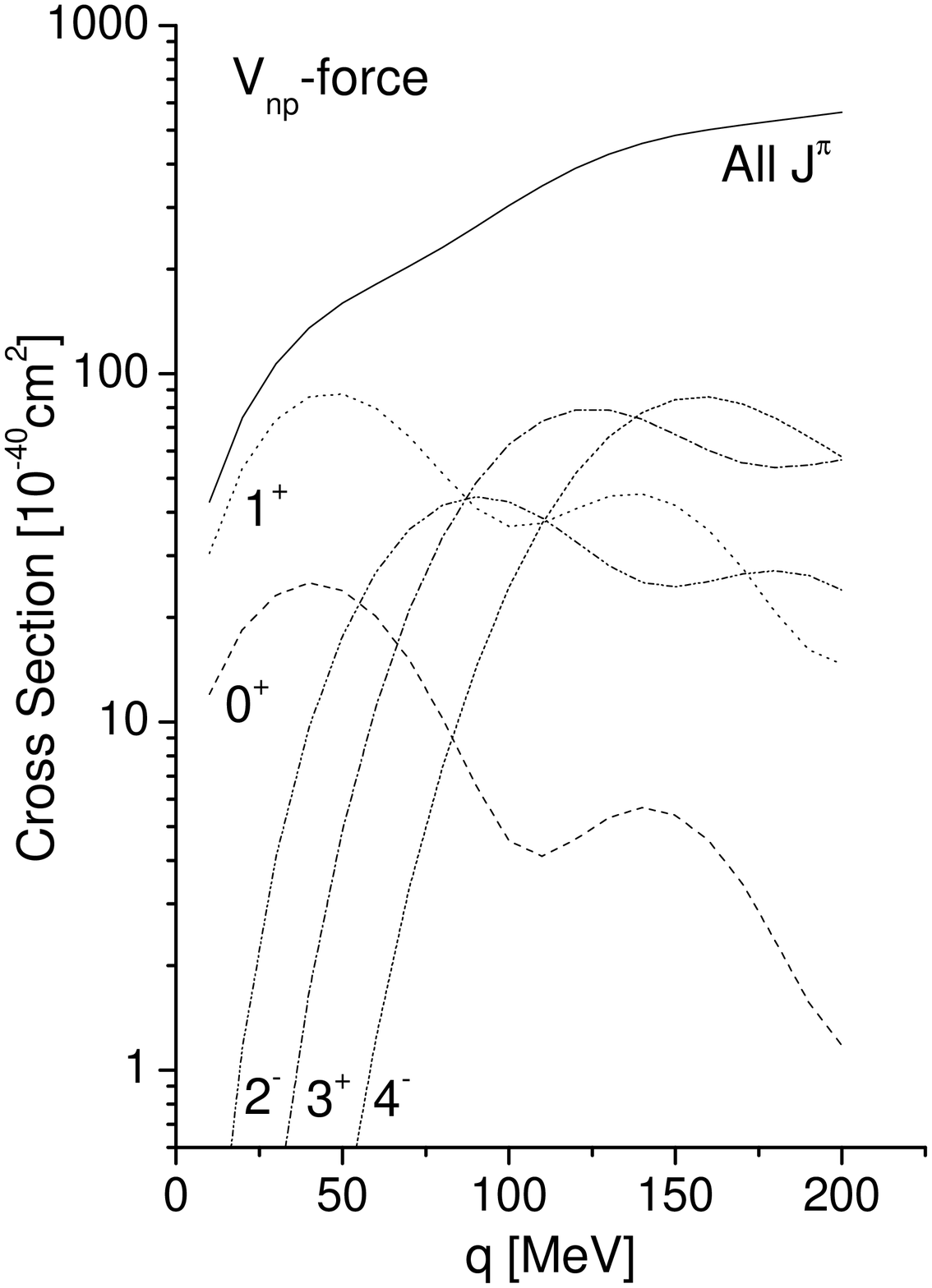}
\caption{ Cross section, as a function of the momentum transferred
by the neutrino. The results corresponding to nuclear structure
calculations of states $J^\pi$ in $^{208}$Bi performed in the
single particle basis which includes only bound and quasi bound
states and using a separable multipole force interaction are
shown. Solid line represents the results obtained by adding up all
states up to $J^\pi \leq 10^{\pm}$, the other curves show the
results of some selected states. \label{fig:fig3}}
\end{figure}

\begin{figure}[h!]
\includegraphics[width=10cm]{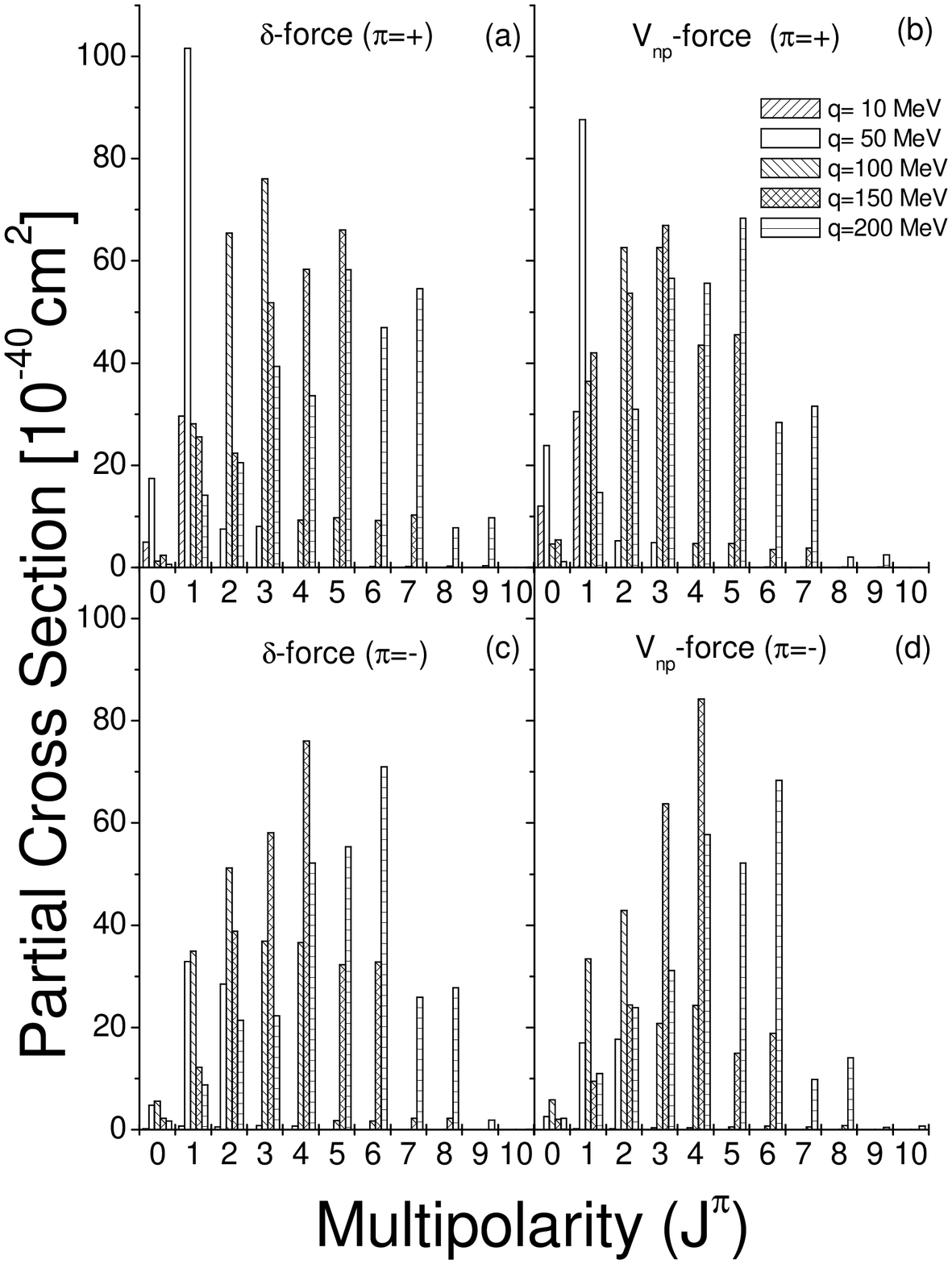}
\caption{ Multipole decomposition of the contrbutions to the cross
section, for some selected values of the momentum $q$. The upper
insets (a) and (b) show the contributions of positive parity states,
lower insets (c) and (d) show the results of negative parity states.
The insets at the left, (a) and (c), show the results obtained by
using a delta force interection, right hand side insets (b) and (d)
show the results obtained with a separable proton-neutron
interaction, as explained in the text . \label{fig:fig4}}
\end{figure}

In order to determine the order of magnitude of the cross section,
an issue which may still be controversial in view of some recently
published results \cite{lang1}, we have performed a detailed
comparison between the nuclear structure calculations described
before and those of \cite{alexa}. The calculations of Ref.
\cite{alexa} are particularly accurate for the description of the
low-energy portion (excitation energies lower than 3-4 MeV) of the
spectrum of $^{208}$Bi. By the other hand, the calculations
performed by using the separable interaction introduced in the
previous paragraphs are well suited for the description of the
higher portion of the spectrum., since the parameters of the
interaction have been adjusted in order to reproduce the position
and intensity distribution of the Isobaric Analogue State (IAS)
and Giant Gamow-Teller Resonance (GTR). However, since little is
known about the energy distribution of other multipole states,
some doubts may arise concerning the reliability of the calculated
wave functions of other multipolarities. Figures (\ref{fig:fig2})
to (\ref{fig:fig4}) show the results which we have obtained by
using both the $\delta$-force interaction of \cite{alexa}, and the
separable multipole-multipole interaction of Eq.(\ref{vsep}). This
set of results correspond to the diagonalization of both
interactions in a single particle basis which includes only bound
and quasi-bound states. The similarity between the results is
undeniable, adding confidence to the present results, which for
the values of the momentum transferred considered agree also with
the results of Volpe et al. \cite{volpe}.(see our
Figs.(\ref{fig:fig2}) and (\ref{fig:fig3}) and Table 1 of
\cite{volpe}, for q=100 MeV, that is $\sigma\approx 4.16 \times
10^{-38}$ cm$^2$ (of Ref. \cite{volpe}) and $\sigma\approx 3.04
\times 10^{-38}$ cm$^2$ (present)).

b)Multipole decomposition

Figure \ref{fig:fig4} shows the contributions of all multipole
states considered in our calculations, for some selected values of
the momentum transferred. The comparison of the results shown in
insets (a)-(d) of this figure indicates that both interactions
yield to very similar results, and that the contributions reach a
maximum at about $q=100-150$ MeV. Both interactions show a
sizeable contributions from the $J^\pi=1^+$ states, among the
positive parity states, and for the $J^\pi=2^-,3^-, 4^-$ states,
for the negative parity states.

c)Effects due to the inclusion of resonant and continuum states

\begin{figure}[h!]
\includegraphics[width=10cm]{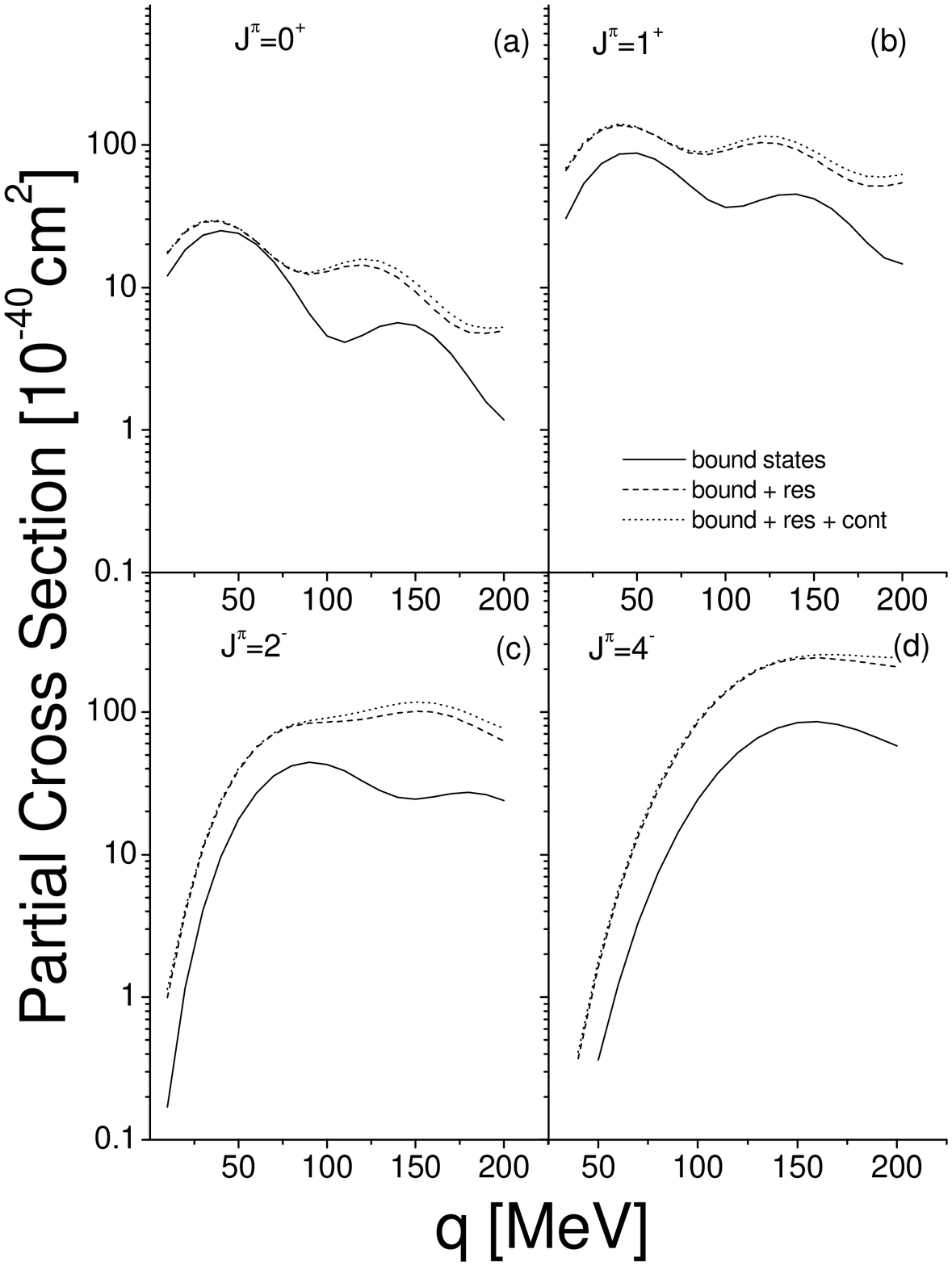}
\caption{Contributions to the cross section, for some selected
multipole states, obtained with the separable multipole-multipole
interaction. Insets (a)-(d) show the results corresponding to
$J^\pi=0^+,1^+,2^-$, and $4^-$ states. Solid lines show the
results of bound and quasi-bound states, dashed and dotted lines
show the results obtained by including resonant states and
scattering (continuum) states \label{fig:fig5}}
\end{figure}

In Figure \ref{fig:fig5} we show the contributions of allowed
Fermi and Gamow-Teller transitions to the cross section of Eq.
(\ref{sigma}). The calculations have been performed by keeping the
different classes of states which define the single-particle
basis. The curves labelled {\it{bound, resonant}}, and
{\it{continuum}} indicate the contribution of configurations, of
the nuclear wave functions, where the proton-particle state is a
bound, resonant or continuum state. The results show a sizeable
contribution of resonant states, which increases with the neutrino
energy, and a negligible contribution from states in the
continuum, in spite of the huge number of continuum states
included. This is, somehow, an expected result, because the time
scale involved in the decay of single-particle resonances is much
larger than the time available for the energy transfer from the
incoming neutrino. However, in Refs. \cite{hayes,hayes-towner}, it
was speculated on that the continuum could play a significant role
in neutrino-nucleus interactions, because of the increase of the
cross section at energies of the order of 100 MeV. This is not
what we have found consistently in all of our results. Also in the
case of the forbidden transitions, the effect due to the inclusion
of the continuum is minor.

\begin{figure}[h!]
\includegraphics[width=10cm]{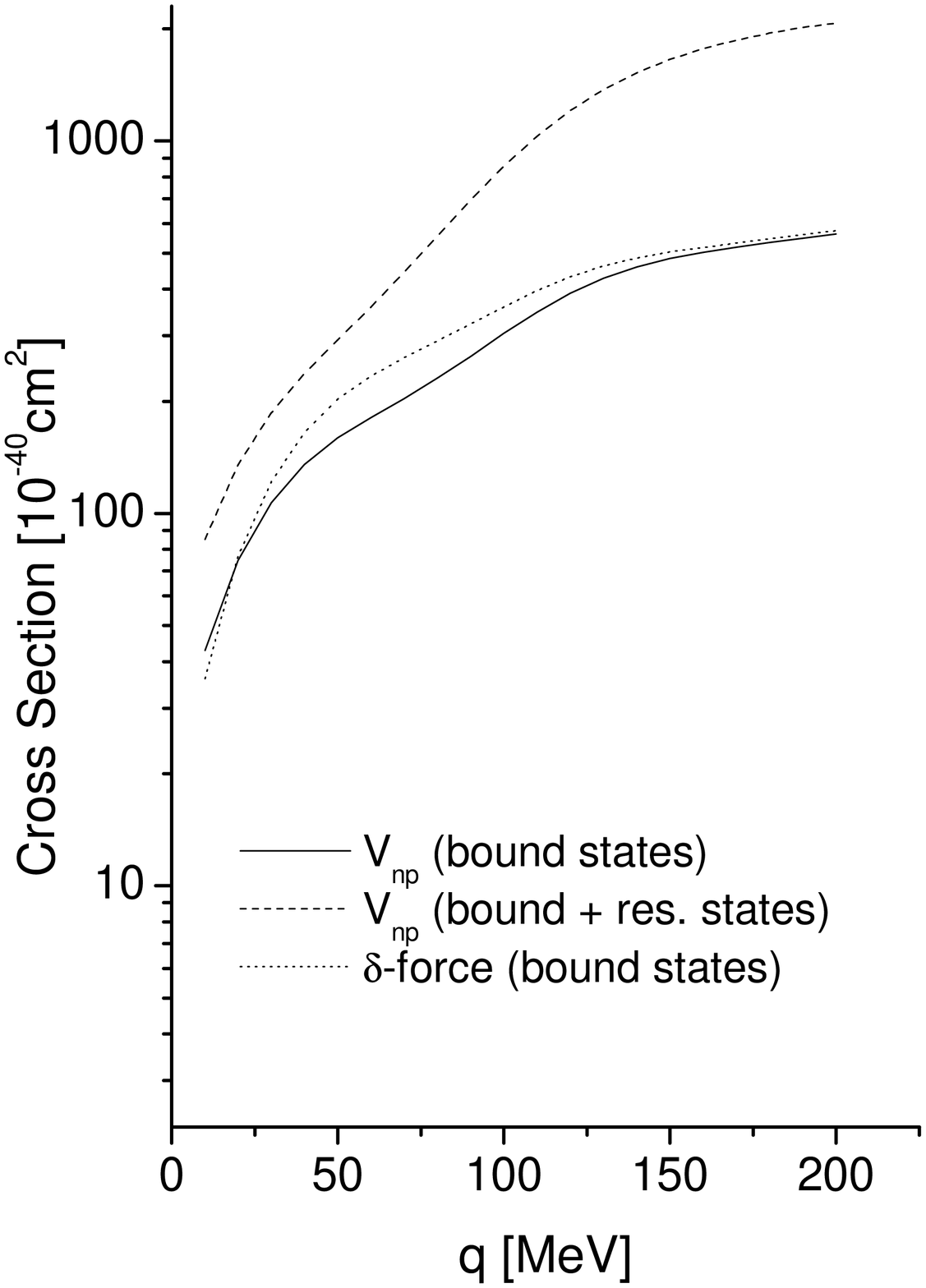}
\caption{Total cross section (sum over all multipoles and
parities), as a function of the momentum $q$, corresponding to the
separable interaction (solid line) and delta force
interaction(dotted lines) in a basis of bound and quasibound
states. The results corresponding to the separable interaction
$V_{np}$, diagonalized in the basis which includes bound,
quasibound and resonant states, are shown with dashed lines.
\label{fig:fig6}}
\end{figure}

Figure \ref{fig:fig6} shows the comparison of the results
corresponding to bound, quasi-bound and resonant states. It
becomes evident that the inclusion of resonant states increases
significantly the values of the cross section, for momentum
transfer higher that 100 MeV. One should notice that in getting
these results we have not restricted the number of resonant states
by keeping, for instance, very narrow resonances only. If one does
it, the increase of the cross section at higher energies is
smaller than the one shows in Figure \ref{fig:fig6}, but still
seizable. A noticeable feature of the curves of Figure
\ref{fig:fig6} is the saturation of the cross section, at values
of the order of $\sigma \approx 6 \times$ $10^{-38}$ cm$^2$ (bound
and quasi-bound states) and $\sigma \approx 18 \times$ $10^{-38}$
cm$^2$ (bound, quasi-bound and resonant states).



As an overall feature, the contribution of the continuum-states is
to be neglected since the bulk of the cross section is given by
bound and resonant states.

\section{Conclusions}\label{conclusions}
In this work we have calculated some of the processes which
contribute to the cross section for neutrino charged-current
interactions on lead. We focus on the $\gamma$ and proton emission
channels from $^{208}$Bi. The nuclear structure part of the
calculation was done by enlarging the single-particle basis to
accommodate for bound, resonant and continuum single-particle
states. In this enlarged basis we have calculated nuclear wave
functions for states belonging to the spectrum of $^{208}$Bi. The
description of the nuclear states was performed by a direct
diagonalization of an effective, separable, interaction with
multipole strengths adjusted to the phenomenology. The quality of
the adjusted interaction was established, also, by a direct
comparison with the force employed by \cite{alexa}. The cross
section was calculated by performing a partial wave expansion of
the lepton wave functions and by computing the matrix elements of
isospin and isospin-spin multipole operators between the ground
state of $^{208}$Pb and excited states of $^{208}$Bi.

The results of the calculations, show that : i) the largest
contributions to the considered channels of the cross section are
given by nuclear excitations where bound and resonant states
participate as proton single-particle states, and ii) the
contribution of single-particle states in the continuum is, for all
practical purposes, negligible.

Although the results may be taken as a single-case sample, since
we have considered just the case of neutrino-electron reaction on
Pb, we think that they are representative of the situation which
may emerge in neutrino reactions on other targets, like $^{12}$C.
Work is in progress concerning this light mass target.

Because of the structure of the nuclear transition operators
involved in the calculations, we expect to find similar results in
the description of single beta-decay processes. Work is in
progress concerning single and double beta decay calculations with
the inclusion of resonant single-particle states.

\section{Acknowledgement}\label{acknowledgement}

This work has been partially supported by the National Research
Council (CONICET) of Argentina. Discussions with Drs S. Elliott
and K. Kubodera are acknowledged with pleasure. We are
particularly indebted to Dr. Cristina Volpe, for her useful
comments about the momentum dependence of the cross section, for
helping in finding errors in an earlier version of the manuscript,
and for her careful reading of the present version.

\appendix

\section{The operators and their matrix elements}
\label{operators}

The wave functions of single-particle states are written as the
product of radial, orbital, and spin wave functions:
\begin{equation}
\Psi_{s.p}=R(r)i^l {\left[Y_{l}(\hat{r})\chi^{1/2}\right]}_{jm} .
\end{equation}
where $R(r)$ is the radial part of the wave function,
$Y_{lm_l}(\hat{r})$ is the orbital component of the angular
momentum and $\chi^{1/2}_{m_s}$ is the spin component. The
operators which enter the definition of the current-current
electroweak interaction are written:
\begin{equation}
T_{(L,\gamma=1)\lambda \mu}=j_L(qr)i^L{\left[Y_L \times
\sigma\right]}_{\lambda \mu} \tau_- ,
\end{equation}
for spin-dependent operators, and
\begin{equation}
T_{(L,\gamma=0)\lambda \mu}=j_L(qr)i^LY _{\lambda=L,
\mu=M_L}\tau_-,
\end{equation}
for spin-independent operators, $j_L(qr)$ is the regular Bessel
function of integer order. The tensor operators can be expressed
in terms of proton(particle)-neutron(hole) configurations as:
\begin{eqnarray}
T_{\lambda \mu}&=&\sum_{pn}<p \mid T_\lambda \mid
{\bar{n}} > a^{\dagger}_{p} a_{\bar{n}} \nonumber \\
&=&\sum_{pn}\frac{<p \mid \mid T_\lambda \mid \mid n>}{2 \lambda +
1}{(a^{\dagger}_{p}b^{\dagger}_{n})}_{\lambda \mu} .
\end{eqnarray}
In this notation the operator $a^\dagger$ creates a proton and the
operator $b^\dagger$ creates a neutron-hole. The reduced matrix
elements of the operators $F_{\lambda \mu}$ are given by the
expression
\begin{equation}
<p\mid\mid T_\lambda \mid\mid n>=F(pn,qL)G(pn,L\gamma \lambda),
\end{equation}
where
\begin{eqnarray}
F(pn,qL)&=&\int dr\; r^2 R_p(r)j_L(qr)R_n(r) ,
\end{eqnarray}
and
\begin{eqnarray}
 G(pn,L\gamma=1,
\lambda)&=&\hat{\lambda}\hat{j}_n\hat{j}_p\hat{L}\hat{l}_n
i^{l_n-l_p+L}<l_n 0 L 0\mid l_p 0> \sqrt{\frac{3}{2\pi}}
{\left\{
\begin{array}{ccc}
l_n & 1/2 & j_n \cr L & 1 & \lambda \cr
l_p & 1/2 & j_p
\end{array}
\right\}},
\end{eqnarray}
are the radial integral and the angular momentum re-coupling
factors for spin-dependent operators $(\gamma=1)$, and
\begin{eqnarray}
F(pn,qL)&=&\int dr\; r^2 R_p(r)j_L(qr)R_n(r),
\end{eqnarray}
and
\begin{eqnarray}
G(pn,L\gamma=0, \lambda=L)&=&\hat{j}_n\hat{j}_p\hat{L}\hat{l}_n
i^{l_n-l_p+L}<l_n 0 L 0\mid l_p 0> \sqrt{\frac{1}{8\pi}}
{(-1)}^{j_n+1/2+l_p+L} {\left\{
\begin{array}{ccc}
l_n & j_n & 1/2
\cr j_p & l_p & L
\end{array}
\right\}},
\end{eqnarray}
for spin-independent operators $(\gamma=0)$, respectively.

In the above equation $\hat{k}=\sqrt{2j_k+1}$, and the adopted
coupling scheme is always $|{(l s)}{jm}>$ and the standard phases
for the angular momentum re-couplings \cite{bmv1}.

\section{Nuclear transition density}
\label{density}

The single particle basis has been constructed by including bound
neutron states and bound, resonant and continuum proton states. We
performed a TDA calculation of the spectrum of $^{208}$Bi, by
diagonalizing the interaction
\begin{eqnarray}\label{inter}
V=\sum_{L \gamma \lambda}g_{L \gamma \lambda}{(T_{L \gamma
\lambda}.T_{L \gamma \lambda})}_0 \label{vsep}
\end{eqnarray}
where the tensor operators are defined by the tensor product of
the orbital and spin operators (see the previous Appendix)
\begin{equation}
T_{L \gamma=1, \lambda \mu}=i^L f_L(r) {(Y_L \times
\sigma)}_{\lambda \mu}
\end{equation}
with $\lambda=0,1,2,....$, $|\lambda-1| \leq L \leq \lambda+1 $
and parity $\pi={(-1)}^L$ and
\begin{eqnarray}
T_{L \gamma=0, \lambda=L\; \mu}=i^L f_L(r) Y_{\lambda \mu}
\end{eqnarray}
for $\lambda=L=0,1,2,...$, and $g_{L \gamma \lambda}$ is the
strength of the interaction in the channel $(L \gamma) \lambda$.
The actual values of $g_{L \gamma \lambda}$ are adjusted to
reproduce the experimental data, either the energy of the
low-lying states or the giant resonances, for each set of
excitations.

The matrix element of a two-body interaction (\ref{inter}),
between particle-hole states may be written in terms of the matrix
elements between two-particle configurations:
\begin{eqnarray}
<pn^{-1}:J|V|p'{n'}^{-1}:J>=-\sum_{J'}(2J'+1) \left\{
\begin{array}{ccc}
p&n&J \\
p'&n'&J'
\end{array}
\right\}<pn':J'|V|p'n:J'>
\end{eqnarray}

Therefore, the matrix element of the interaction in the particle
representation is of the form
\begin{eqnarray}
<pn':J'|V|p'n:J'> &=&\sum_{L \gamma \lambda}g_{L \gamma \lambda}
\left\{
\begin{array}{ccc}
p'&n&J' \\
n'&p&\lambda
\end{array}
\right\}{(2\lambda+1)}^{-1/2}{(-1)}^{\lambda+p'+n'+J'}\nonumber\\
& &<p||T_{L \gamma \lambda}||p'><n'||T_{L \gamma\lambda}||n>
\end{eqnarray}
and, by writing the spin scalar ($\gamma=0$) and spin vector
($\gamma=1$) tensor components explicitly one has:

\begin{eqnarray}
<pn':J'|V|p'n:J'> &=&\sum_{L \gamma \lambda}g_{L \gamma \lambda}
\left\{
\begin{array}{ccc}
p'&n&J' \\
n'&p&\lambda
\end{array}
\right\}{(2\lambda+1)}^{-1/2}{(-1)}^{\lambda+p'+n'+J'}\nonumber\\
& &(\delta_{\gamma,0}\delta_{L,\lambda}<p||i^L f_L(r) Y_L
||p'><n'||i^L f_L(r) Y_L||n>\nonumber \\
& &+\delta_{\gamma,1}\delta_{L,\lambda \pm 1}<p||i^L f_L(r) {(Y_L
\times \sigma)}_{\lambda}||p'><n'||i^L f_L(r) {(Y_L \times
\sigma)}_{\lambda}||n>)\nonumber \\
\end{eqnarray}

The determination of the amplitudes $C^{(k)}(pn,J^\pi)$ leads to
the calculation of the transition amplitudes
\begin{equation}
\rho(pn;0 \rightarrow {J}_f^\pi,k)=\delta_{\lambda
J_f}C^{(k)}(pn,J^\pi)
\end{equation}
Then, with these transition amplitudes, the matrix elements needed
to calculate the neutrino-nucleus cross section are written
\begin{equation}
M_{(L \gamma)
\lambda}(q)=\sum_{pn}\frac{1}{\sqrt{2\lambda+1}}F(pn,qL)G(pn,L\gamma,\lambda)
\rho(pn;0 \rightarrow {J}_f^\pi,k)
\end{equation}
and they are functions of the lepton momentum transfer $q$.
\section{Form factors}\label{formfactors}
To obtain the factors $\lambda_\tau$ and $\lambda_{\sigma \tau}$
of Eq.(\ref{mnuc}) we expand the leptonic and hadronic currents of
the effective weak Hamiltonian, in terms of the momentum transfer
$q$. By keeping contributions up to the order $\frac{1}{M_N}$,
where $M_N$ is the nucleon mass, the effective, minimal, weak
Hamiltonian is written
\begin{eqnarray}\label{hweak}
H_{\rm{weak}}&=&-\frac{G_F}{\sqrt{2}}\cos{\theta_{\rm{C}}}\left[J^\mu
j_{\mu}\right]
\end{eqnarray}
where $j_{\mu}$ is the lepton current
\begin{equation}
j_{\mu}=\bar{\psi}_e\gamma_\mu(1+\gamma_5)\psi_\nu
\end{equation}
and $J^{\mu}$ is the nucleon current in the limit of low momentum,
with components
\begin{eqnarray}
J^0&=&f_V(q)-\frac{1}{2M}f_A(q)(\sigma \cdot \mathbf {q})\nonumber
\\
\mathbf{J}&=&-i f_A(q)\sigma+ \frac{1}{2M_N}(f_V(q)-2M_Nf_W(q))
(\sigma \times \mathbf{q})+i f_V(q) \frac{\mathbf{q}}{2M_N}
\end{eqnarray}
After some straightforward algebra, the calculation of the matrix
elements of Eq.(\ref{cross}) yields
\begin{eqnarray}\label{formfactor}
\lambda_\tau &=& f^2_V(q)(1+\cos{\theta}) \left[ 1+2 \left(
\frac{E_e-E_\nu}{2M_N} \right)\right] \nonumber \\
\lambda_{\sigma \tau}&=&f^2_A(q)\left[
\left(1-\frac{1}{3}\cos{\theta}\right)+\frac{2}{3} \left(
\frac{E_e-E_\nu}{2M_N} \right)(1+\cos{\theta})-\frac{4}{3}
 \left(
\frac{E_e+E_\nu}{2M_N}
\right)\left(\frac{f_V(q)-2M_Nf_W(q)}{f_A(q)}\right)(1-\cos{\theta})
\right],\nonumber \\
\lambda_\Lambda&=&4 f_A^2(q)\left[\cos\theta+
\left(\frac{E_l-E_\nu}{2M_N}\right) \left(1+\cos\theta\right)
+\left(\frac{E_l+E_\nu}{2M_N}\right)
 \left(\frac{f_V(q)-2 M_Nf_W(q)}{f_A(q)}\right)
 \left(1-\cos\theta\right)\right] \, \, \, .
\end{eqnarray}
which agrees with the results of \cite{fukugita,kuramoto}. The
nucleon form-factors, for the axial(A), vector(V), and
weak-magnetic(W) terms, $f_{V,A,W}$, have the following momentum
dependence \cite{kuramoto}
\begin{eqnarray}
f_A(q)&=&-\frac{1.262}{{\left( 1+ \frac{q^2}{{(1.032
\rm{GeV})}^2}\right)}^2}\nonumber \\
f_V(q)&=&\frac{1}{{\left( 1+ \frac{q^2}{{(0.84
\rm{GeV})}^2}\right)}^2}\nonumber \\
f_{W}(q)&=&-\frac{3.706}{2M_N}f_V(q)
\end{eqnarray}
In the above equations we have neglected pseudo-scalar terms,
because they are smaller than the included terms by factors of the
order $\frac{1}{M_N}$.


\begin{thebibliography}{99}
\bibitem{ref1} A. B. Balantekin and G. M. Fuller, J. Phys. G {\bf 29}, 2513 (2003)
\bibitem{ref2} A. B. Balantekin and H. Yuksel, New J.Phys. {\bf 7}, 51 (2005)
\bibitem{ref3} Y. Z. Qian et al, Phys. Rev. Lett. {\bf 71}, 1965 (1993)
\bibitem{sno} Q. R. Ahmad et al., Phys. Rev. Lett. {\bf 89}, 011301 (2002)
\bibitem{sk} Y. Fukuda et al., Phys. Rev. Lett. {\bf 81}, 1562 (1998)
\bibitem{ref5} J. Bahcall, M. H. Pinsonneault and S. Basu, Astrophys. J. {\bf 555}, 990 (2001)
\bibitem{ref6} J. Bahcall and C. Pe\~na-Garay, New J. Phys. {\bf 6}, 63 (2004)
\bibitem{dbd} H. Ejiri, Phys. Rep. {\bf 338}, 265 (2000).
\bibitem{nastro} J. D. Vergados, Phys. Rep. {\bf 361}, 1 (2002).
\bibitem{fuller} G. M. Fuller, W. C. Haxton and G. C. McLaughlin, Phys. Rev. D {\bf 59}, 085005 (1999)
\bibitem{elliot} S. R. Elliot, Phys. Rev. C {\bf 62}, 065802 (2000)
\bibitem{noble} A. J. Noble; Proceedings of the Symposium {\it Physics
in Collision}, Annecy, France, June 26-29, 2007.
\bibitem{engel} J. Engel, G. C. McLaughlin and C. Volpe, Phys. Rev. D {\bf 67}, 013005 (2003)
\bibitem{fukugita} M. Fukugita, Y. Kohyama, K. Kubodera and T. Kuramoto, Astrophys. J. {\bf 337}, 59 (1989)
\bibitem{kuramoto} T. Kuramoto, M. Fukugita, Y. Kohyama and K. Kubodera, Nucl. Phys. A {\bf 512}, 711 (1990)
\bibitem{volpe1} C. Volpe, J. Phys. G {\bf
30}, L1-L6 (2004); ibid hep-ph/0303222.
\bibitem{bmvII}A. Bohr and B. Mottelson, Nuclear Structure vol. 2, Benjamin Readings,
M. A, (1975).
\bibitem{hayes} A. C. Hayes. APS Meeting Abstracts, 202 (1997).
\bibitem{hayes-towner}A. C. Hayes and I. S. Towner, Phys. Rev. C {\bf
61}, 044603 (2000).
\bibitem{kolbe}E. Kolbe and K. Langanke, Phys. Rev. C. {\bf 63},
025802 (2001).
\bibitem{langanke}E. Kolbe, K. Langanke and P. Vogel, Phys. Rev. D. {\bf 66},
013007 (2002).
\bibitem{ian}J. Blomqvist, O. Civitarese, E. Kirchuk, R. J. Liotta, T. Vertse,
Phys. Rev. C {\bf 53} (1996) 2001.
\bibitem{cg} O. Civitarese and M. Gadella, Phys. Rep. {\bf 396},
41 (2004).
\bibitem{vol03}
A. Volya and V. Zelevinsky, Phys. Rev. C {\bf 67}, 054322  (2003)
and references therein.
\bibitem{id02}
R. Id Betan, R. J. Liotta, N. Sandulescu and T. Vertse, Phys. Rev.
Lett. {\bf 89}, 042501 (2002).
\bibitem{ver88}
T. Vertse, P. Curutchet, O. Civitarese, L. S. Ferreira and R. J.
Liotta, Phys. Rev. C {\bf 37}, 876  (1988).
\bibitem{report} J. Suhonen and O. Civitarese, Phys. Rep. {\bf 300}, 123 (1998)
\bibitem{b68}
T. Berggren, Nucl. Phys. {\bf A 109}, 265 (1968).
\bibitem{lio96}
R. J. Liotta, E. Maglione, N. Sandulescu and T. Vertse, Phys.
Lett. {\bf B 367}, 1 (1996).
\bibitem{id05}
R. Id Betan, R. J. Liotta, N. Sandulescu, T. Vertse and R. Wyss,
Phys. Rev. {\bf C 72}, 054322 (2005).
\bibitem{bl}
T. Berggren and P. Lind, Phys. Rev. {\bf C 47}, 768 (1993).
\bibitem{vlm}
T. Vertse, R. J. Liotta and E. Maglione, Nucl. Phys. {\bf A 584},
13 (1995).
\bibitem{cvl}P. Curutchet, T. Vertse and R. J. Liotta,
Phys. Rev. {\bf C 39}, 1020 (1989).
\bibitem{ixa}L. Gr. Ixaru, M. Rizea and T. Vertse,
Comput. Phys. Commun. {\bf 85}, 217 (1995).
\bibitem{alexa}
P. Alexa, J. Kvasil and R. Sheline, Phys. Rev. C {\bf 55}, 3170
(1997).
\bibitem{Ber78}
T. Berggren, Phys. Lett. {\bf B 73}, 389 (1978).
\bibitem{lang1}K.Langanke et al., Phys. Rev. Lett. {\bf 100},
011101 (2008).
\bibitem{volpe}R. Lazauskas and C. Volpe, Nucl. Phys. {\bf A 792},
219 (2007).
\bibitem{bmv1}A. Bohr and B. Mottelson, Nuclear Structure vol. I, W. A. Benjamin Inc.,
M. A, (1975).
\end{thebibliography}
\end{document}